\documentclass[runningheads]{llncs}
\usepackage[T1]{fontenc}
\usepackage{graphicx}
\usepackage{xspace}
\usepackage{amsmath,amsfonts}
\usepackage[linesnumbered,ruled,vlined]{algorithm2e}
\usepackage{array}
\usepackage{textcomp}
\usepackage{stfloats}
\usepackage{url}
\usepackage{verbatim}
\usepackage{graphicx}
\usepackage{cite}
\usepackage{amsmath,amsfonts}
\usepackage{graphicx}
\usepackage{hyperref}
\usepackage{textcomp}
\usepackage{xcolor}
\usepackage{amssymb}
\usepackage{booktabs}
\usepackage{makecell}
\usepackage{listings}
\usepackage[most]{tcolorbox}
\newcommand{\dm}{\textsc{DeepMutation}\xspace}
\newcommand{\dc}{\textsc{DeepCrime}\xspace}

\newboolean{showcomments}
\setboolean{showcomments}{true}
\ifthenelse{\boolean{showcomments}}
  {\newcommand{\nb}[2]{
  \fbox{\bfseries\sffamily\scriptsize#1}
     {\sf\small$\blacktriangleright$\textit{\textcolor{red}{#2}}$\blacktriangleleft$}
   }
  }
  {\newcommand{\nb}[2]{}
   
  }

\newcommand{\COMMENT}[1]{}

\begin{document}
\title{New Formulation of DNN Statistical Mutation Killing for Ensuring Monotonicity: A Technical Report}
\titlerunning{New Formulation of DNN Statistical Mutation Killing}
% If the paper title is too long for the running head, you can set
% an abbreviated paper title here
%
% \author{Jinhan Kim\inst{1} \and
% Second Author\inst{1}\orcidID{1111-2222-3333-4444} \and
% Third Author\inst{3}\orcidID{2222--3333-4444-5555}}
%

% First names are abbreviated in the running head.
% If there are more than two authors, 'et al.' is used.
%
% \institute{Princeton University, Princeton NJ 08544, USA \and
% Springer Heidelberg, Tiergartenstr. 17, 69121 Heidelberg, Germany
% \email{lncs@springer.com}\\
% \url{http://www.springer.com/gp/computer-science/lncs} \and
% ABC Institute, Rupert-Karls-University Heidelberg, Heidelberg, Germany\\
% \email{\{abc,lncs\}@uni-heidelberg.de}}

\author{
Jinhan Kim\inst{1} \and
Nargiz Humbatova\inst{1} \and
Gunel Jahangirova\inst{2} \and
Shin Yoo\inst{3} \and
Paolo Tonella\inst{1}
}

\institute{
Università della Svizzera italiana, Lugano, Switzerland \\
\email{\{jinhan.kim, nargiz.humbatova, paolo.tonella\}@usi.ch}
\and
King's College London, London, UK \\
\email{gunel.jahangirova@kcl.ac.uk}
\and
KAIST, Daejeon, Republic of Korea \\
\email{shin.yoo@kaist.ac.kr}
}

\authorrunning{Kim et al.}
\maketitle              % typeset the header of the contribution
\begin{abstract}
Mutation testing has emerged as a powerful technique for evaluating the effectiveness of test suites for Deep Neural Networks (DNNs). Among existing approaches, DeepCrime's statistical mutant killing criterion has leveraged statistical testing to determine whether a mutant significantly differs from the original model. However, it suffers from a critical limitation: it violates the monotonicity property, meaning that expanding a test set may result in previously killed mutants no longer being classified as killed. In this technical report, we propose a new formulation of statistical mutant killing based on Fisher’s exact test that preserves the statistical rigour of it while ensuring monotonicity.

\end{abstract}
\section{Introduction}
% \han{We need to be clear about what the instances are. Consult muff paper later.}
% \dmm, \dm, \dc, \muff
% \han{Comment}

Mutation testing involves introducing small, artificial changes, referred to as mutations, into a system and assessing whether a given test set can detect these changes~\cite{papadakis2019mutation}. In traditional software systems, mutation testing has been extensively studied and is known to be an effective means of evaluating test adequacy. However, adapting this concept to DNNs presents unique challenges due to their non-deterministic nature, continuous output behaviour, and dependence on the training process and data.

Recent studies have proposed various mutation operators tailored specifically for DNNs, which can be broadly categorised into pre-training~\cite{humbatova2021deepcrime} and post-training mutation operators~\cite{deepmut, deepmut++, kim2025muff}. These operators simulate different types of faults, ranging from data corruption in the training set to weight perturbations in the trained model. Alongside these operators, several definitions of mutant killing and mutation score have been introduced. A notable example is the statistical killing definition introduced by Jahangirova \& Tonella~\cite{jahangirovantonella}, which was later implemented in a mutation testing tool DeepCrime~\cite{humbatova2021deepcrime}. This approach applies statistical hypothesis testing to assess whether a mutant behaves significantly differently from the original model, accounting for the non-determinism inherent in DNN training.

Despite its strengths, our analysis identifies a limitation of this approach: it violates the \textit{monotonicity} property. In the context of test adequacy, monotonicity requires that if a mutant is killed by a given test set, it should also be killed by any superset of that test set. However, we observe that under the existing statistical killing definition, expanding the test set can cause a previously killed mutant to be deemed as \textit{not killed}. This non-monotonic behaviour is counterintuitive and undesirable, as it undermines the reliability and consistency expected from a robust test adequacy criterion.

In this technical report, we propose a new formulation of statistical mutant killing that preserves the statistical rigour of the statistical killing definition by Jahangirova \& Tonella~\cite{jahangirovantonella} while addressing its monotonicity issue. Our approach leverages Fisher’s exact test~\cite{agresti1992survey} to assess whether individual inputs contribute to distinguishing between the original model and its mutants. By aggregating these contributions across the test set, we ensure that adding more test cases cannot reduce the killing status of a mutant, thereby guaranteeing monotonicity.

\section{Background}

Mutation analysis is a fault-based testing technique traditionally used to evaluate the adequacy of test suites~\cite{papadakis2019mutation}. It involves introducing artificial faults (mutants) into a system and measuring the effectiveness of a test set by its ability to `kill' these mutants, i.e., detect deviations from the original behaviour. In the context of DNNs, mutation analysis has been adapted by defining specialised mutation operators (MOs) suitable for neural architectures.

\subsection{Mutation Operators for DNNs}

Mutation operators for DNNs are generally classified into two categories:

\begin{itemize}
    \item Pre-training MOs~\cite{humbatova2021deepcrime}: These modify the training set, the model structure, or hyperparameters before model training, requiring training for each mutant.
    \item Post-training MOs~\cite{deepmut, deepmut++, kim2025muff}: These directly alter the weights, biases, and model structure of the trained model without retraining.
\end{itemize}

Post-training MOs, while not necessarily representative of real faults, offer substantial computational efficiency and have been shown to exhibit strong correlation with fault detection capability~\cite{abbasishahkoo2024teasma}. Moreover, they have been found to be more sensitive to the strength of the test set than pre-training MOs~\cite{kim2025muff}.

\subsection{Mutation Killing Definitions Introduced So Far}

Abbasishahkoo et al.~\cite{abbasishahkoo2024teasma} have analysed various Mutation Score (MS) formulations proposed for DNN mutation analysis. In this section, we revisit those approaches with an emphasis on their underlying definitions of mutant killing.

Let $N$ denote the original DNN model, $M = \{M_1, M_2, \ldots, M_n\}$ the set of generated mutants, and $T$ a test set. The following killing definitions have been introduced:

\begin{itemize}

    \item \textbf{\dc's Statistical Killing (KD1)~\cite{jahangirovantonella, humbatova2021deepcrime}}: When mutants are instantiated multiple times, to account for the stochastic nature of training, each model (original or mutant) yields a distribution of accuracy values (or other performance metrics) for a given test set $T$. Let $A_N = \{\text{acc}_{N_1}, \ldots, \text{acc}_{N_r}\}$ and $A_{M_i} = \{\text{acc}_{M_{i,1}}, \ldots, \text{acc}_{M_{i,r}}\}$ be the sets of accuracies from $r$ original model and mutant instances. A mutant $M_i$ is considered killed if the difference in accuracy is statistically significant:
    % \nargiz{I think performance metric would be more accurate because it concerns not only classification systems}
% \nargiz{I think, we always considered that r should be = s. I am also not sure that this can be credited to DeepCrime, given that it was proposed by Jahangirova and Tonella}

    \[
    \text{killed}(T, M_i) =
    \begin{cases}
    1, & \text{if } p\text{-value} < \alpha \text{ and } \text{effectSize} \ge \beta \\
    0, & \text{otherwise}
    \end{cases}
    \]
    
    \noindent where the $p$-value is computed from a statistical test comparing $A_N$ and $A_{M_i}$, and the effect size is measured using Cohen's $d$.

    \item \textbf{\dm's Standard Killing (KD2)~\cite{deepmut}}: A mutant $M_i$ is considered killed by $T$ if there exists at least one input $t \in T$ such that $N(t)$ is correct and $M_i(t)$ is incorrect. The killing condition is:

    \[
    \text{killed}(T, M_i) = 
    \begin{cases}
    1, & \text{if } \exists t \in T \text{ such that } N(t) \text{ is correct and } M_i(t) \text{ is incorrect} \\
    0, & \text{otherwise}
    \end{cases}
    \]

    \item \textbf{\dm's Class-Level Killing (KD3)~\cite{deepmut}}: For classification tasks with $k$ classes $C = \{C_1, C_2, \ldots, C_k\}$, a mutant $M_i$ is considered killed on class $C_j$ by input $t$ $\in C_j$  if $N(t) = C_j$ and $M_i(t) \ne C_j$. 

    \[
    \text{killedClass}(t, M_i, C_j) = 
    \begin{cases}
    1, & \text{if } N(t) = C_j \text{ and } M_i(t) \ne C_j \\
    0, & \text{otherwise}
    \end{cases}
    \]

\noindent This definition can be extended to a test set $T$ by evaluating all inputs in $T$. A test set $T$ is said to kill mutant $M_i$ on class $C_j$ if there exists at least one input $t \in T$ such that $\text{killedClass}(t, M_i, C_j) = 1$:

    \[
    \text{killedClass}(T, M_i, C_j) = 
    \begin{cases}
    1, & \text{if } \exists t \in T \text{ such that } \text{killedClass}(t, M_i, C_j) = 1 \\
    0, & \text{otherwise}
    \end{cases}
    \]

    Recently, Abbasishahkoo et al.~\cite{abbasishahkoo2024teasma} found that the mutation score based on KD3 shows the strongest correlation with fault detection rate, suggesting it is the most effective metric for evaluating test adequacy in DNNs, where the faults under consideration are defined as groups of similar mispredicted inputs.
 % \nargiz{If I recall correctly, in this work, the authors have quite a different notion of fault than we do. they call a fault - a cluster of inputs}    
    \item \textbf{\dm's Input-Level Killing (KD4)~\cite{deepmut++}}: At the input level, a mutant $M_i$ is considered killed by input $t$ if it produces a different output than the original model $N$: %\nargiz{Wouldn't it be better if this one comes first, before KD1? Otherwise we could specify that this one comes from DeepMutation++, not DeepMutation}

    \[
    \text{killed}(t, M_i) = 
    \begin{cases}
    1, & \text{if } M_i(t) \ne N(t) \\
    0, & \text{otherwise}
    \end{cases}
    \]

\end{itemize}

% \nargiz{there is also a threshold-based killing that was analysed by Jahangirova and Tonella}

\subsection{Remarks}

Although various kill definitions have been proposed, no single formulation has gained widespread adoption in the literature. KD2 is often disregarded as it is generally a too lenient criterion (i.e., a criterion that makes any mutant easy-to-kill). Both KD2 and KD3 lack statistical rigour, meaning that the non-determinism of the training process is not accounted for and a mutant might be killed when considering a model instance, while it is not killed if another instance is obtained from the training process.
Despite its solid statistical basis, one reason for the lack of consensus on KD1 (\dc's Statistical Killing) is that it relies on the availability of multiple instances of both the original and mutated models to support a statistical comparison, which makes it computationally expensive. Moreover, it is a common practice with post-training mutation to mutate a single model instance, disregarding the non-deterministic effects of the training process.

In this technical report, we focus on a critical issue that affects KD1, despite its statistical foundations: it violates monotonicity when the test set is augmented. The other definitions (KD2, KD3, and KD4) do not suffer from this problem. In the following section, we present a detailed analysis of this issue with reference to KD1.

\section{KD1's Violation of Monotonicity}

Let us consider a scenario where the test set $T$ has already killed the pre-training mutant $M_i$, meaning that the accuracy distributions $A_N$ and $A_{M_i}$, obtained from multiple instances of the original model $N$ and mutant $M_i$, are statistically significantly different according to the criteria of KD1. 

Now, let us suppose we augment the test set by adding a new set of tests $S$ to form $T' = T \cup S$, and subsequently measure the new accuracy distributions $A'_N$ and $A'_{M_i}$ on this enlarged test set. Intuitively, one would expect that since $M_i$ was killed by $T$, it should also be killed by any superset $T'$ of $T$. This property is known as \emph{monotonicity} of the kill definition with respect to test set inclusion. However, in the case of KD1, there is no guarantee that a statistically significant difference between $A'_N$ and $A'_{M_i}$ will hold. In fact, it is possible for the addition of tests in $S$ to dilute the observed difference in accuracies, causing the $p$-value to increase above the significance threshold $\alpha$, or the effect size to drop below the required threshold $\beta$. This can happen if the new tests in $S$ are easy for the mutant $M_i$ (hence, both original's and mutant's outputs are correct) or  challenging for the original model $N$ (hence, both original's and mutant's outputs are incorrect), thereby reducing the relative performance gap. 

Consequently, the mutant $M_i$ that was previously considered killed by $T$ may no longer be considered killed by the larger test set $T'$, violating monotonicity. This counter-intuitive behaviour undermines the reliability of KD1 as a test adequacy metric, especially when it is used to guide test set augmentation. In fact, increasing the test set should not invalidate the previously observed killed capability. %, while other kill definitions such as KD1, KD2, and KD3 are based on input-level or class-level comparisons that preserve monotonicity by construction; if a mutant is killed by some input or class, adding more tests cannot revert that kill status.

To empirically expose this violation, we conducted a case study using the MNIST handwritten digit recognition dataset~\cite{mnist} and the HBS mutation operator of \dc. Our experimental setup involved evaluating 20 instances of both the original model and the mutated model across cumulative test sets of increasing size, ranging from 100 to 10k tests in increments of 100.

For each test set size, we computed the accuracy distributions $A_N$ and $A_{M_i}$ and applied KD1's statistical significance criteria: a two-sample t-test with significance level $\alpha = 0.05$ and effect size threshold $\beta = 0.2$ (Cohen's d). A mutant was considered killed if the p-value was below $\alpha$ and the original model's mean accuracy exceeded the mutant's mean accuracy.

\begin{figure}[t]
\centering
    \includegraphics[width=0.8\columnwidth]{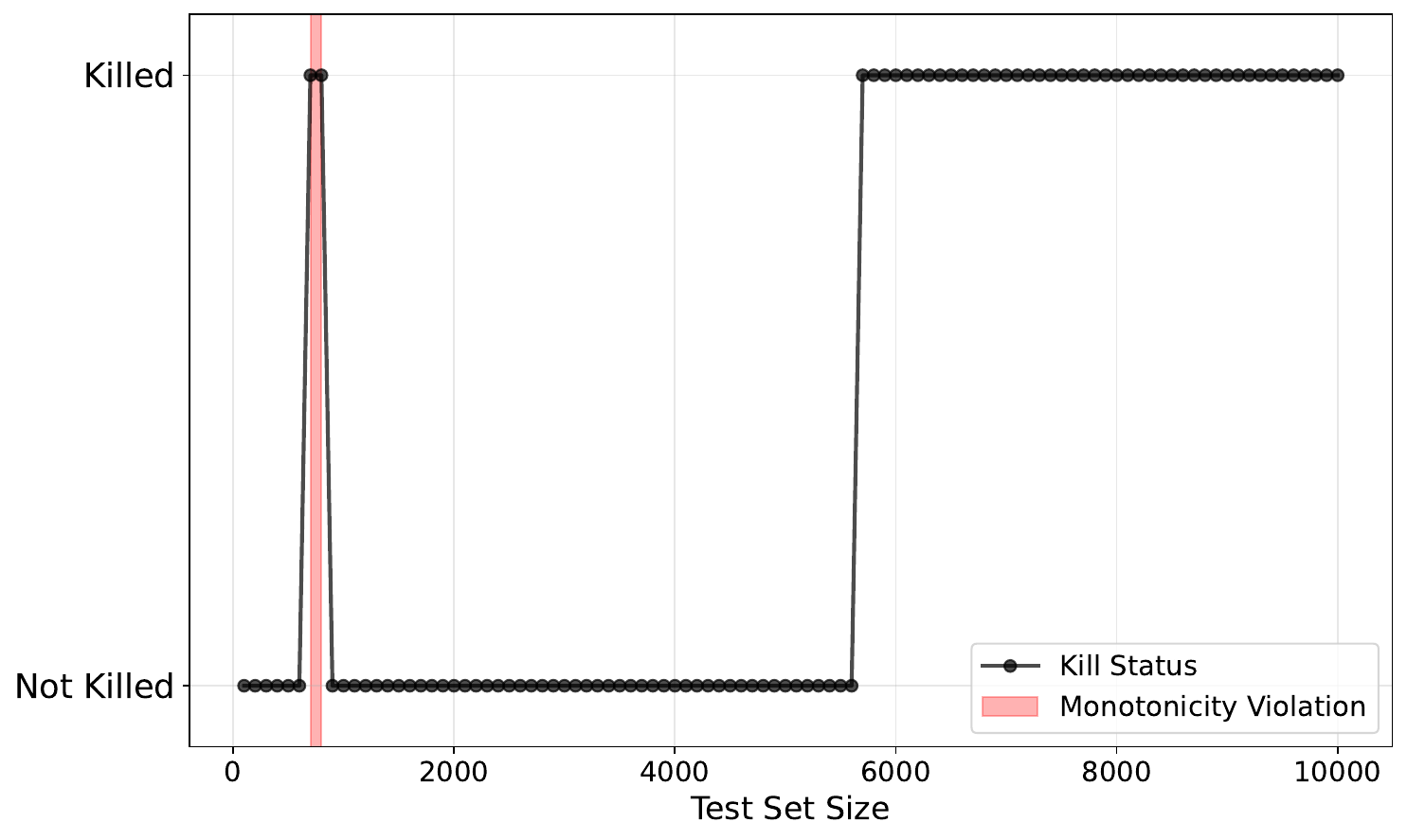}
    \caption{Kill status of mutant across test set sizes. Red shaded regions indicate monotonicity violations where mutants are killed at smaller test sizes but not killed at larger test sizes.}
    \label{fig:kill_status}
\end{figure}

Figure~\ref{fig:kill_status} presents the kill status of the HBS mutant across different test set sizes. It reveals clear evidence of monotonicity violations, with two instances of consecutive augmented test sets with the kill status transitioning from `Not Killed' to `Killed' as the test set size increases, while it later drops back to `Not Killed' when more tests are added. Specifically, we identified four cases of mutants that were killed at test size 800 but not at test size 900. This empirical evidence confirms our theoretical analysis, showing that KD1 can violate the monotonicity property.

\section{New Formulation of DNN Statistical Mutation Killing}

The goal of this new formulation is twofold: (a) to retain the assumption by Jahangirova \& Tonella~\cite{jahangirovantonella} that mutation killing should be defined statistically based on multiple model instances, in order to account for the non-deterministic nature of DNN training; and (b) to address the issue of monotonicity violations, where a superset of test inputs fails to kill a mutant while a subset succeeds.

\begin{table}[h!]
\centering
\caption{Example 2×2 Contingency Table}\label{tab:cont_table}
\begin{tabular}{|l|c|c|}
\hline
 & Correct & Incorrect \\
\hline
Original Insts. & 17 & 3 \\
\hline
Mutant Insts. & 11 & 9 \\
\hline
\end{tabular}
\end{table}

To achieve this, we utilise Fisher’s exact test~\cite{agresti1992survey} because it enables us to define mutant killing at the level of individual test inputs, and then naturally extend this definition to a test set by measuring the proportion of mutants killed by the test set. This formulation inherently preserves monotonicity as adding non-killing inputs cannot decrease the overall proportion of killed mutants (i.e., the killing inputs remain such, regardless of the addition of non-killing ones). At the same time, we maintain the statistical ground of KD1, by considering multiple instances of both original and mutated models.

Given an input and a mutant, we construct a 2×2 contingency table as shown in Table~\ref{tab:cont_table}. In this table, rows represent original versus mutated model instances, and columns represent whether the prediction on a given input was correct or incorrect. Let $a$, $b$, $c$, and $d$ denote the cell counts in the contingency table:

\[
\begin{bmatrix}
a & b \\
c & d \\
\end{bmatrix}
=
\begin{bmatrix}
\text{Correct}_{\text{orig}} & \text{Incorrect}_{\text{orig}} \\
\text{Correct}_{\text{mut}} & \text{Incorrect}_{\text{mut}} \\
\end{bmatrix}
=
\begin{bmatrix}
17 & 3 \\
11 & 9 \\
\end{bmatrix}
\]

Fisher’s exact test calculates the likelihood of observing a difference in predictions between original and mutated models as strong as the one shown, assuming that both models perform the same (the null hypothesis is that there is no difference between original and mutant predictions). The $p$-value is given by:

\[
p = \frac{ \binom{a + b}{a} \binom{c + d}{c} }{ \binom{a + b + c + d}{a + c} }
\]

For example, Table~\ref{tab:cont_table} presents such a table based on 20 instances of the original and mutated models. Applying Fisher’s exact test to this table yields a $p$-value of 0.082, which exceeds the commonly used significance threshold of 0.05. This suggests no statistically significant difference in behaviour between the original and mutated models on this input. However, if we increase the number of correct predictions by the original model from 17 to 18 (and reduce incorrect predictions from 3 to 2), the resulting $p$-value drops to 0.031, which is below 0.05, indicating a statistically significant difference. This shows that Fisher’s exact test can determine whether an input distinguishes the original from the mutated model in a statistically meaningful way. We refer to such inputs as \textit{killing inputs}. Formally, an input $x$ kills a mutant $M$ if the resulting contingency table yields a $p$-value below a pre-defined threshold $\alpha$ (commonly 0.05). A mutant is considered \textit{killed} by a test set if at least one input in the set kills it. The overall mutation score is defined as the proportion of mutants killed by the test set.

In our new definition of mutant killing, the presence of one test input that kills the mutant is enough to consider the mutant as killed. While this preserves monotonicity, it results in a coarse-grained mutation score that may fail to differentiate between test sets of varying strength. To address this limitation, we introduce a secondary metric: NKI (Number of Killing Inputs), defined as the number of inputs in a test set $T$ that kill a given mutant $M$ according to Fisher’s exact test. It is still monotonic, while providing a finer-grain indication of the strength of each test set.  When two test sets $T_1$ and $T_2$ achieve the same mutation score, we can compare them by the NKI. It can also be noticed that NKI can be computed for a subset of all mutants, or even for individual mutants: if some mutants are known to be particularly challenging (e.g., because they are associated with a small number of killing inputs), we consider only those mutants when using NKI to compare $T_1$ to $T_2$. 
We can now redefine mutation killing as NKI $\geq$ 1 for a given mutant $M$, which can be generalised to NKI $\geq \tau$: users may adopt a stronger criterion for killing a mutant $M$ than the existence of a single killing input, by requiring at least $\tau$, a user-defined threshold. 
Usage of NKI, possibly on the subset of most challenging mutants, is expected to provide higher sensitivity than the mutation score when comparing two test sets.

% While this definition applies to individual inputs, it can extend to input sets. For a given test set $T = \{t_1, t_2, \dots, t_n\}$, we construct a 2×2 contingency table for each input $t_i$ and apply Fisher’s exact test to determine if $t_i$ is a contributing input. We define the contributing input set as:
% \[
% C_T (M_i) = \{t_i \in T \mid p(t_i) < \alpha\}
% \]
% \noindent where $\alpha$ is the significance level (commonly $\alpha = 0.05$), and $p(t_i)$ is the $p$-value from Fisher's test for input $t_i$. To determine whether a test set $T$ kills a mutant, we compare the proportion of contributing inputs:
% \[
% \phi(T, M_i) = \frac{|C_T (M_i)|}{|T|}
% \]
% We then define a test set $T$ as killing a mutant $M_i$ if:
% \[
% \text{killed}(T, M_i) = 
% \begin{cases}
% 1, & \text{if } \phi(T, M_i) \ge \tau \\
% 0, & \text{otherwise}
% \end{cases}
% \]
% \noindent where $\tau$ is a predefined threshold. To make threshold selection easier, we propose using the training set as a reference. Specifically, the proportion of contributing inputs in the training set can serve as a baseline. For instance, if 3\% of training inputs are contributing, then the test set must also contain at least 3\% (i.e., $\tau$) contributing inputs to be considered effective in killing the mutant.

Our new formulation achieves both goals stated above. First, by applying Fisher’s test to correctness outcomes collected from multiple original and mutated model instances, it preserves the core statistical perspective introduced by Jahangirova \& Tonella~\cite{jahangirovantonella}, thereby accommodating the non-deterministic nature of DNN training. Second, by evaluating mutant killing for each input independently, it naturally avoids monotonicity violations: adding non-killing inputs to a test set cannot make a mutant not killed any more, if at least one killing input already exists in the initial test set, ensuring that larger test sets do not paradoxically appear less effective than their subsets.
To ensure high sensitivity and discriminative power of our new mutation killing criterion, we introduced a secondary metric, NKI, which can be used to evaluate the relative strength of alternative test sets that achieve the same or similar mutation scores. Overall, the new mutation killing criterion and the new mutation killing metrics are statistically grounded, monotonic and sensitive to the different killing capabilities of alternative test sets.

% For instance, if $3\%$ of training inputs are contributing, then a test set must also contain at least $3\%$ contributing inputs to be considered effective in killing the mutant.

% Odds ratio: 4.636363636363637
% P-value: 0.08235877701951116

% original_error_count: The number of inputs from the training set that the original model misclassified.

% contributing_inputs: The number of inputs for which the difference between the original and mutant models is statistically significant. (i.e., Inputs that pass the fisher test of the contingency

% Kill Definition: # of contributing inputs > original _error_count • Remove mutants, if the number of contributing inputs in the training set does not exceed the original model's error count.

% \section{Experiment}

% \section{Remarks on Defining Mutation Score}

% For KD1, its MS is very similar to what traditional mutation testing's MS.
% For KD2, as it is defined over the classes, its MS is defined over the number of total mutants multiplied by the number of classes.
% For KD3, the MS is defined over the total number of \textit{tests} in a test set.
% for KD4, the MS is defined over the mutants killed by train set 

\section{Conclusion}

In this technical report, we have identified and addressed a critical issue with DeepCrime's statistical mutant killing criterion (KD1) in DNN mutation testing: its violation of the monotonicity property. By introducing a new formulation based on Fisher’s exact test that operates at the level of each individual input, we ensure that the killed status of a mutant never degrades when the test set is expanded. Our approach retains the statistical foundation of KD1 by considering multiple model instances, thereby accounting for the non-deterministic nature of DNN training, while addressing its theoretical shortcomings. Future work includes empirical evaluation of the proposed criterion and metrics across diverse DNN architectures and datasets, and exploration of its effectiveness for different tasks, such as test input generation, prioritization and selection. We conjecture high sensitivity of our new metrics based on the theoretical analysis, but of course an empirical validation of such conjecture would be quite important.

\bibliographystyle{splncs04}
\bibliography{bib}

\end{document}